\documentclass[aps,twocolumn,preprintnumbers,superscriptaddress,amsmath,amssymb,showpacs]{revtex4}
\usepackage{graphicx}% Include files
\usepackage{dcolumn}% Align table columns on decimal point
\usepackage{bm}% bold math
\begin{document}

\title{Geometric Theory of Columnar Phases on Curved Substrates}
\author{Christian D. Santangelo}
\author{Vincenzo Vitelli}
\author{Randall D. Kamien}
%\email{kamien@physics.upenn.edu}
\affiliation{
Department of Physics and Astronomy, University of Pennsylvania, Philadelphia PA, 19104}
\author{David R. Nelson}
\affiliation{Department of Physics, Harvard University, Cambridge MA, 02138}
\date{\today}

\begin{abstract}
We study thin self-assembled columns constrained to lie on a curved, rigid substrate.  The curvature presents no {\sl local} obstruction to equally spaced columns in contrast to curved crystals for which the crystalline bonds are frustrated. Instead, the vanishing compressional strain of the columns implies that their normals lie on geodesics which converge (diverge) in regions of positive (negative) Gaussian curvature, in analogy to the focussing of light rays by a lens. We show that the out of plane bending of the cylinders acts as an effective ordering field. 
\end{abstract}

%\keywords{Geometric frustration, curvature, smectic, columnar, copolymers.}
\maketitle

The concept of long range order is the lynchpin of much of our understanding of both phase transitions and the elasticity of the resulting crystalline and liquid crystalline phases.  In the presence of a quenched perturbation, this order can be disrupted \cite{Larkin, Imry}.  Moreover, in low dimensional systems thermal fluctuations can destroy long-range order: in two dimensions long range order cannot be achieved in systems with continuous symmetry \cite{Mermin}, while in two and three dimensions one-dimensional periodic order is unstable to thermal fluctuations \cite{Landau}.  It would thus seem unlikely that there would be much to say about one-dimensional periodic order on a two-dimensional substrate.  However, most of these ``no-go'' results suppose short range interactions.  When the substrate is curved, however, long-range interactions arising from global geometric constraints can lead to ordering where none is usually expected.  This feature opens up a novel venue for directing the self-assembly of thin layers of material with a desired structure by appropriately engineering the underlying substrate. Experimental realizations include self-assembled monolayers of tilted molecules \cite{stellacci}, nematic double emulsions \cite{Alberto06} and solid rafts on curved interfaces \cite{baus03}. 
The presence of geometric constraints causes strong elastic deformations from the ground state of these materials that  often cannot be captured by perturbative methods. In many cases, the investigation of these challenging issues of non-linear elasticity can benefit from blending classical continuum mechanics with modern geometrical methods \cite{Cerda-Mahadevan,Chris-Randy,Me-Ari}. The deformations of thin elastic shells is one such problems that occupies a central place in any exposition of the basic principles of applied mechanics, but still continues to stimulate a wealth of theoretical and experimental research in analogous systems broadly defined as curved solids \cite{bowi2000,vitelucks06}. 
 
In this article, we address the less understood case of  thin films composed of nearly incompressible columns such as those obtained from the self-assembly of block copolymers in a cylindrical phase \cite{Bate99}. This system presents a natural arena to study how the curvature of the substrate affects the translational and  orientational order that arises by minimizing the distortion energy of the cylinders \cite{Harrison}. Cylindrical domains of block copolymers also have technological potential as nanowires and templates for quantum dots \cite{Park-Chaikin97}.  Non-planar confining geometries, for example cylinders with radius comparable but incommensurate to the copolymer period, have been successfully employed to generate structures not accessible under planar confinement \cite{Russell04}. In the opposite limit, films of block copolymer cylinders (with tens of nm radius)  have been synthesized on substrates made of larger, melted colloidal particles (with few $\mu$m radius) \cite{Alex06}.  There is markedly increased orientational order of the cylinders on these curved substrates when compared with flat substrates.  Here we propose a coupling mechanism 
between the local normal to the columns and the varying curvature of the substrate that acts as an effective ordering field for the columnar order.  In particular, we argue that where the Gaussian curvature of the substrate changes sign a boundary-like term, akin to a weak anchoring term in a liquid crystal cell, aligns the columns.  The constraint of equal spacing then dictates the order in the region of negative Gaussian curvature.

The free energy density for columns has two essential ingredients: the compression energy which sets the intercolumnar spacing, and the bending energy measured by the three dimensional curvature, $\kappa$, of the columns.  The curvature is further decomposed into the geodesic curvature, $\kappa_g$, and the normal curvature, $\kappa_n$:  We denote the normal to the surface as $\bf{N}$, the unit tangent vector of the columns as ${\bf t}$, and define ${\bf n}\equiv \bf{N}\times{\bf t}$ to be the normal to the columns in the tangent plane of the surface.  Then $\kappa^2=\vert\partial_s{\bf t}\vert^2=\vert(\bf{t}\cdot\nabla){\bf t}\vert^2$, $\kappa_g= \bf{n}\cdot({\bf t}\cdot\nabla)\bf{t}$ and $\kappa_n=\bf{N}\cdot(\bf{t}\cdot\nabla)\bf{t}$ \cite{foot1}, from which it follows that $\kappa^2=\kappa_g^2+\kappa_n^2$.  The normal curvature measures the bending of the cylinder as it is constrained to lie on the surface whereas the geodesic curvature measures the intrinsic curvature of the cylinder bending within the surface.  For example, great circles on a sphere have vanishing geodesic curvature yet nonzero normal curvature because they still bend.

The compression energy density is most easily expressed in terms of a phase field $\Phi({\bf x})$, which is related to the density $\rho =\rho_0 + \rho_1\cos(2\pi\Phi/a)$ of one of the polymeric blocks.  Here, $a$ is the preferred intercolumn spacing, $\rho_0$ is the constant background density, and $\rho_1$ is the order parameter for one-dimensional crystalline order.  In this language, the columns sit at the level sets of $\Phi = m a$ ($m\in\mathbb{Z}$) and the columns are at their equilibrium spacing when $\vert\nabla \Phi\vert=1$.  The compressional strain, $e$, must be written in terms of $\vert \nabla \Phi \vert$ and must vanish when $\vert\nabla\Phi\vert=1$; for example, 
$e=\frac{1}{2}\left[1-(\nabla\Phi)^2\right]$ is one of the many acceptable forms.  The precise form of $e$ is unimportant for our discussion, as we shall see.
The total free energy in the columnar phase is
\begin{equation}
F=\frac{B}{2}\int d^2\!x\,\left[e^2 + \lambda^2\kappa^2\right] \!\!\!\! \quad ,
\end{equation}
where $\lambda\equiv \sqrt{K_1/B}$ is the penetration depth, $K_1$ is  the splay modulus, and $B$ is the bulk modulus.  In the experiments of Ref.~\cite{Alex06} $\lambda\approx3~{\rm nm}$ is a tenth of the cylinder spacing.  It is useful to note that the derivative of $\vert\nabla\Phi\vert={\bf n}\cdot\nabla\Phi$ perpendicular to the columns is $P^j_{i}\partial_j\vert\nabla\Phi\vert = \vert\nabla\Phi\vert ({\bf n}\cdot{\bf D})n_i$ where
$P^{j}_{i}=\delta^j_{i}-n^j n_i$ is the transverse projection operator and ${\bf D}$ is the covariant derivative in the surface. If the strain vanishes, then $\vert\nabla\Phi\vert=1$ everywhere and
 $(\bf{n}\cdot{\bf D}){\bf n}=0$. We recognize this as the geodesic equation for $\bf n$, and conclude that vanishing strain implies that the normals to the columns are geodesics on the surface.
\begin{figure}[t]
\includegraphics[width=2.2truein]{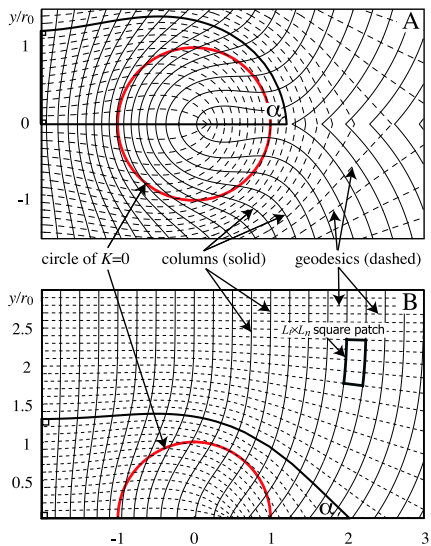}
\caption{\label{fig:focusing} Configuration for columns (solid curves) shown in projection from above on a surface with height function $h(x,y) = h_0 \exp [-(x^2+y^2)/(2 r_0^2)]$.  We set the column at $x=-100 r_0$ to be straight and the red circle is the line with $K=0$.  The column normals lie on geodesics (dashed lines) that diverge and converge in regions of negative and positive Gaussian curvature, respectively.  The dark lines delineate a triangle made of three geodesics: the $x$-axis, the straight column at $x=-\infty$, and the geodesic that bends towards the
$x$-axis.  In A) we show the full bump for $h_{0}=4r_0$. Note that some geodesics turn by $\pi$.   In B) we depict  the region $y>0$ for a bump with $h_{0}=2r_0$. For clarity, we stop the geodesics when they intersect other geodesics. }
\vskip-0.2truein
\end{figure}

This observation suggests an analogy between this system and geometric optics, where the column normals take the place of light rays.  The Gaussian curvature $K$ of the surface plays the role of the index of refraction in focussing the geodesics, as is evident from the geodesic deviation equation, $\partial^2 \xi/\partial s^2 = - K \xi$, where $s$ is the arclength along the geodesics and $\xi$ is a measure (precisely, the Jacobi field which points from one geodesic to another \cite{DoCarmo-book}) of the distance between two infinitesimally-close and parallel geodesics.  Thus, $K<0$ causes the column normals to diverge while $K>0$ causes them to converge.  We illustrate this in Fig. \ref{fig:focusing} for columns aligned by a smooth wall far to the left of a Gaussian-shaped bump, which sets up a family of initially parallel geodesics. The figure shows the effect of geometrical frustration on column-by-column growth that proceeds from left to right. The bending of the geodesics leads to sharp kinks in the columns downstream from the bump similar to the formation of caustics in ray optics.   Indeed, this analogy between geometry and optics is the basis for gravitational lensing \cite{lensing} and the columnar system provides a ``down to earth'' realization of
this important observational tool.  Note also that geodesics can turn by $\pi$ and return to the column on the left if the bump has a large enough aspect ratio. 

The angle $\alpha$ which describes the downstream grain boundary where two families of columns with  different orientations meet can be related to the Gaussian curvature of the surface by utilizing the Gauss-Bonnet theorem on a surface patch $M$ \cite{DoCarmo-book}:
\begin{equation}\label{eq:GB}
\oint_{\partial M} \kappa_g ds = 2\pi - \sum_i\Delta\theta_i - \int_M K dA \!\!\!\! \quad ,
\end{equation}
where $\Delta\theta_i$ are the external angles at any vertices in $\partial M$, and the integral around $\partial M$ is broken into piecewise integrals along the smooth part of $\partial M$.
For the geodesic triangle depicted in Fig. \ref{fig:focusing}, this calculation yields
\begin{equation}
\alpha = \int_{M} K dA \!\!\!\! \quad ,
\end{equation}
since the three edges are geodesics, two of the external angles are $\pi/2$ and the remaining
external angle is $\pi-\alpha$.
We take the left edge of the triangle out to $x \rightarrow -\infty$.  Notice that as the distance from the center of the bump increases, the triangle covers a larger portion of the surface, asymptotically containing up to half of the bump.  We conclude, therefore, that $\alpha$ approaches zero with increasing $x$ since the integrated Gaussian curvature of any smooth bump tends to zero in this limit.  We expect that near
the kinks, the equal-spacing constraint will relax so as to lower the overall energetics of the caustic structure.  Whether dislocation pairs unbind along the $y$-axis in Fig. \ref{fig:focusing} to soften the curvature or 
whether other low energy configurations \cite{Chris-Randy} which share the same topology come into play is an open question.
In contrast to the case of crystals on curved substrates \cite{bowi2000,vitelucks06}, there is no local obstruction to growing equally spaced columns.  Given \textit{any} single column, the equal-spacing condition $|\nabla \Phi| = 1$ can be integrated to determine the shape of the remaining columns within a finite size region.  The compression strain does not, therefore, impose any local orientational order on the columns.

Since at every point on the surface there are geodesics in every direction, one might think that the  columns could lie along geodesics.  However, this assumption is incompatible with equal spacing on a curved substrate. To see this, assume that the columns lie along geodesics and that
they are equally spaced.  We start with one geodesic column.  At two points on the column $L_t$ apart,
we move off along the normal direction and trace the geodesics obeying $(\bf{n}\cdot{\bf D}){\bf n}=0$.  After integrating both a fixed distance $L_n$, these geodesics will end on a new
column and ${\bf n}$ will still be perpendicular to the column's tangent vector.
Here again, the Gauss-Bonnet theorem (Eq. \ref{eq:GB}) constrains possible column configurations.
In the case of our
$L_t\times L_n$ rectangle (shown in Fig. \ref{fig:focusing}), there are four vertices with $\Delta\theta=\pi/2$ and so 
\begin{equation}\label{eq:GB1}
\oint_{\partial M} \kappa_g ds = - \int_M K dA \!\!\!\! \quad .
\end{equation}
However, we assumed that $\kappa_g=0$ along the normal flows and along the columns.  We see
this is only possible for arbitrary rectangles if $K\equiv 0$ on the substrate!  We have thus established the incompatibility of equal spacing and vanishing geodesic curvature of the columns.  The true ground state will be some compromise between a non-ideal spacing and some curvature of the columns (the configuration shown in Fig. \ref{fig:focusing} depends on the growth history and is unlikely to be the actual ground state). However, we note that it is the gradient of the compression that is proportional to the curvature of the normals and so, dimensionally, the compression energy will grow with an extra
factor of $L^2$ (in a region of size $L$) compared to the bending energy.  We thus expect, at long distance scales, that the curvature of the columns will absorb the effect of the Gaussian curvature and that the column spacing will be set at the ground state value, {\sl i.e.} $e=0$.  
%%%%%%%%%%%%%%%%%%%%%%%%%%%%%%%
\begin{figure}
\includegraphics[width=2.2truein]{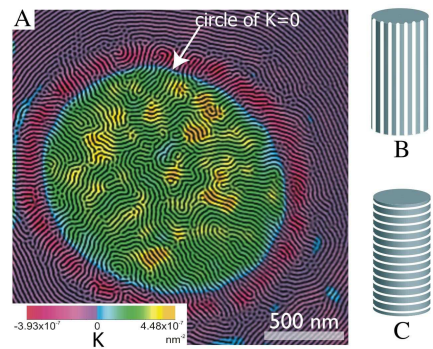}
\caption{\label{fig:layer} A) Data from \cite{Alex06}.  Dark lines are the columns.  The central region has $K>0$, and the outer region has $K<0$.  B) + C) Equally-spaced columns lying on geodesics of a cylindrical surface, a surface with $K=0$.  In B) the columns are straight in three-dimensions and have $\kappa_n=0$, while in C) the columns are curved in three dimensions and $\kappa_n\ne 0$.  Near the circle of $K=0$, the bump shares the geometry of the cylinders.}\vskip-0.2truein
\end{figure}
%%%%%%%%%%%%%%%%%%%%%%%%%%%%%%%

Based on this discussion, it would be natural to suppose that, since the lines of longitude are geodesics on a rotationally symmetric bump, the columns lie along lines of latitude.  However, the curvature of the columns frustrates this expectation.  The intrinsic curvature does not give us a local way of selecting one configuration over another -- one must compute the total curvature energy for different configurations with $e=0$ (Including defects in the ground state, along the lines of Ref. \cite{vitelucks06} is beyond the scope of this paper).  We turn, therefore, to the extrinsic curvature of the columns, $\kappa_n = {\bf N}\cdot({\bf t}\cdot\nabla){\bf t} = -t_i t_j \partial_j N_i$, where the final equality follows from ${\bf N}\cdot{\bf t}=0$.  If we express ${\bf t}=\cos\theta {\bf e}_1 +\sin\theta {\bf e}_2$ in terms of the principal directions ${\bf e}_i$, the extrinsic curvature becomes $\kappa_n=\kappa_1\cos^2\theta +\kappa_2 \sin^2\theta$ where $\kappa_i$ are the principal curvatures \cite{DoCarmo-book}.  
We take ${\bf e}_1$ to be along the radial direction and ${\bf e}_2$ along the azimuthal direction. The minimum of $\kappa_n^2$ depends on whether $K=\kappa_1\kappa_2$ is positive or negative: when $K<0$, the two principal curvatures differ in sign and $\kappa_n^2$ attains its minimum when $\kappa_n$ vanishes, {\sl i.e.} $\tan^2\theta_0 = -\kappa_1/\kappa_2$.
Deviations $\delta=\theta-\theta_0$ from the preferred angle have an energy per unit area of $\Delta f=-2B \lambda^2 K\delta^2 +{\cal O}(\delta^4)$.

For simplicity, we consider a substrate which is a surface of revolution with height function $h(r)$.  Our arguments do not depend on the particular shape of the bump but rather on the generic fact that near the top $K>0$ and at the bottom, in the ``skirt'', $K<0$. Since we want to consider smooth, azimuthally symmetric bumps, we know that $h'(r)$ vanishes at $r=0$ and $r=\infty$ and so somewhere in between, at $r=r_0$, $h''(r_0)=0$.  Since the radial direction is also a principal direction, the principal curvature $\kappa_r$ vanishes there, and subsequently $K=0$  on that circle.  On this circle, we have a universal, preferred direction for the columns -- they lie along the radial direction since $\tan^2\theta_0=0$.  Thus the circle at radius $r=r_0$ imparts a boundary condition, albeit a very soft one.  Near $r=r_0$ the
restoring energy is quartic in $\delta$, not quadratic. A radial alignment of the columns appears to be the energetically favored solution on the ring $r=r_{0}$ for sufficiently steep bumps.  This picture is roughly borne out by experimental observations reproduced in Fig. \ref{fig:layer}  which shows a highly disordered columnar arrangement near the top of the bump, a radial corona of aligned columns near the ring $r=r_{0}$, with a dramatic transition to columnar rings at slightly larger radius. 

Focussing now on the top of a bump, where $K>0$ and $r<r_0$, 
the minimum of $\kappa_n^2$ occurs when $\sin 2\theta_0=0$.  If $\kappa_2>\kappa_1$ then the minimum is at $\theta_0=0$ with a deviation energy of $\Delta f \sim \frac{B}{2} \lambda^2 (\kappa_2-\kappa_1)\kappa_1 \delta^2$, while when $\kappa_2<\kappa_1$, the preferred angle is $\theta_0=\pi/2$ with a deviation energy of $\Delta f \sim  \frac{B}{2} \lambda^2 (\kappa_1-\kappa_2)\kappa_2 \delta^2$.   Since $\kappa_1$ grows from zero near $r_0$, the radial ordering will still be favored near the band of vanishing $K$.  However, depending on the details of the bump, there will either be azimuthal or radial ordering near the top.   The incompatibility of radial ordering in the $K>0$ region would force dislocations and disclinations into the texture since radial aligned columns have $\kappa_g=0$ and so, by our
original arguments, equal spacing would be impossible.  It may be difficult, however, to distinguish between a slightly disordered azimuthal ordering ($\kappa_1>\kappa_2$) and a dislocation texture arising from boundary conditions at $r=r_0$ when $\kappa_2>\kappa_1$. In contrast, the columns 
on the flanks of the bump are normal to radial geodesics as we may have expected from our initial discussion.  This extrinsic effect should be contrasted to the intrinsic coupling between the column direction and gradients of intrinsic quantities considered in \cite{flexo}.

As stated above, the data of Hexemer {\sl et al.}~\cite{Alex06} finds azimuthal ordering in the $K<0$ regions of the bump with less pronounced order in the $K>0$ regions.   Moreover, in that experiment
there is no long range order in the flat portions of the sample where the restoring force proportional to $K$ vanishes. The translational correlation length between cylinders does not exceed $200 {\rm nm}$, corresponding to a patch of less than ten equidistant columns. On larger distances, the system appears to be comprised of orientationally uncorrelated patches even if a configuration of parallel cylinders is the obvious ground state for perfectly annealed flat samples.  These results are consistent with prior experiments on flat substrates \cite{Hamm05}.

Though the asymptotic shape of the bump controls the asymptotic lines of vanishing normal curvature, the rigidity of equal spacing in the $K<0$ region and the indefatigable radial ordering close to
the inflection point provides (in the absence of defects) a mechanism that generates involutes along the surface \cite{DoCarmo-book}. This follows from the discussion of the geodesic curvature: the violation of equal spacing is more energetically costly than deviations from the preferred curvature. In the experiments \cite{Alex06}, an array of bumps were used as a substrate instead of an isolated bump, as we have considered here.  In this situation the hexagonal geometry implicitly generates more boundaries which act as ordering fields which are likely to promote columnar ordering  in the skirts.  The texture of bumps acts as a source for large core disclinations in the normal field \cite{Me-Ari}.  Our analysis considers a single layer of columns where the bending along their normal direction was neglected.  The experiments imaged the top layer of  a film roughly ten layers thick which was grown on the curved substrate.  Unimaged dislocations and strains are necessary below the surface to accommodate the deformation of the columnar crystal and are  formed as the layer grows on the curved surface.  How the underlying layers couple to the two-dimensional geometry of the topmost columns is an open question.

In summary, we have studied how the curvature of a substrate influences the order of incompressible, periodic structures by drawing an analogy to geometrical optics.  Regions of positive Gaussian curvature cause the convergence of the column normals, negative Gaussian curvature lead to their divergence.  On the one hand, the \textit{extrinsic} curvature of the columns results in a local, preferred ordering depending on the shape of the surface.  On the other hand, on a symmetric bump the region near $K=0$ provides a generic mechanism of radial alignment akin to a soft boundary condition.  We hope to discuss the role of the breakdown of the equal spacing condition and the resulting dislocations in future work.

We thank G. Grason and T.C. Lubensky for discussions. We are grateful to A. Hexemer and E. Kramer for a critical reading of the manuscript and for allowing us to reproduce their experimental data.  CDS, VV, and RDK were supported though the Penn MRSEC Grant DMR-0520020, the Donors of the Petroleum Research Fund, and a gift from L. J. Bernstein. Work by DRN was supported by the Harvard MRSEC Grant DMR-0213805.


\begin{thebibliography}{99}
\bibitem{foot1} In terms of the curvature tensor $K_{ij}$, $\kappa_n = K_{ij}t^i t^j$ where $t_i$ is the tangent to the columns.  Though we could include additional terms \cite{flexo} to model capillarity and wetting effects, such as $K_{ij}n^in^j =\kappa_n -K^i_i$ or $K_{ij}t^in^j$, we neglect them here to emphasize the column-like elasticity.
\end{thebibliography}
\end{document}